# Localization and mass spectrum of $q$-form fields on branes

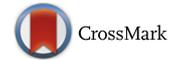

Chun-E Fu [a,b], Yuan Zhong [a], Qun-Ying Xie [c], Yu-Xiao Liu [d,*]

[a] *School of Science, Xi'an Jiaotong University, Xi'an 710049, PR China*
[b] *Department of Physics, University of Toronto, M5S1A7 Toronto, Canada*
[c] *School of Information Science and Engineering, Lanzhou University, Lanzhou 730000, PR China*
[d] *Institute of Theoretical Physics, Lanzhou University, Lanzhou 730000, PR China*

a b s t r a c t

In this paper, we investigate localization of a bulk massless $q$-form field on codimension-one branes by using a new Kaluza–Klein (KK) decomposition, for which there are two types of KK modes for the bulk $q$-form field, the $q$-form and $(q-1)$-form modes. The first modes may be massive or massless while the second ones are all massless. These two types of KK modes satisfy two Schrödinger-like equations. For a five-dimensional brane model with a finite extra dimension, the spectrum of a bulk 3-form field on the brane consists of some massive bound 3-form KK modes as well as some massless bound 2-form ones with different configuration along the extra dimension. These 2-form modes are different from those obtained from a bulk 2-form field. For a five-dimensional degenerated Bloch brane model with an infinite extra dimension, some massive 3-form resonant KK modes and corresponding massless 2-form resonant ones are obtained for a bulk 3-form field.

## 1. Introduction

In the 1920s, Kaluza and Klein presented a unified field theory of gravitation and electromagnetism with the idea of a fifth dimension (one compact extra dimension) beyond the usual four of space and time. The Kaluza–Klein (KK) theory was a purely classical extension of general relativity to five dimensions. Several decades later, the idea that the space–time may be more than four dimensions has attracted increasingly more attention, especially when the braneworld models were proposed [1–9], which were motivated by the classical problems such as cosmological and hierarchy problems. Extra dimensions are usually assumed to be compacted in order to not conflict with our visible four-dimensional world. However, they may also be infinite if the brane has a tension and the matter fields or particles are confined on the brane, which is the idea of the Randall–Sundrum (RS) thin brane scenario [5]. In this thin brane scenario, the infinite extra dimensions are invisible for a four-dimensional observer on the brane and the four-dimensional Newtonian potential can be recovered at large distances.

Subsequently, it was found that the RS thin brane can be generalized to more realistic thick branes generated dynamically by scalar fields and/or some other matter fields [10–28], or by pure gravity [29,30]. In the thick brane scenario, one should not simply assume that the matter fields are trapped on the brane since the energy density of the brane smoothly distributed along the extra dimensions; therefore, it is important and interesting to investigate localization mechanisms for various bulk matter fields [31–55]. On the other hand, investigations of the Kaluza–Klein (KK) modes of bulk fields can provide a way to probe extra dimensions. For example, the massive KK modes for a metric and a vector field can give corrections to Newton's law and Coulomb's law, respectively (the zero modes for them are responsible for the well-known Newton's law and Coulomb's law at low energy) [5,56–58]. So it is necessary to get the mass spectra and eigenfunctions of these bulk fields through a localization mechanism.

In this paper, we focus on the localization of a massless $q$-form field on a codimension-one $p$-brane. The action of the $q$-form field $X_{[q]}$ in the bulk is written as [37]

$$S = -\frac{1}{2(q+1)!} \int d^D x \sqrt{-g}\, Y^{M_1 M_2 \cdots M_{q+1}} Y_{M_1 M_2 \cdots M_{q+1}}, \qquad (1)$$

where the field strength $Y_{[q+1]}$ is defined as $Y_{[q+1]} = dX_{[q]}$ or $Y_{M_1 M_2 \ldots M_{q+1}} = \partial_{[M_1} X_{M_2 \ldots M_{q+1}]}$. The 0-form and 1-form fields are just scalar and vector fields. The higher $q$-form fields with $q \geq 2$ are new types of fields in a higher-dimensional space–time with dimension $D > 4$, which are used to solve some unknown problems such as the cosmological constant problem or dark energy

\* Corresponding author.
*E-mail address:* liuyx@lzu.edu.cn (Y.-X. Liu).

http://dx.doi.org/10.1016/j.physletb.2016.03.069
0370-2693/© 2016 The Authors. Published by Elsevier B.V. This is an open access article under the CC BY license (http://creativecommons.org/licenses/by/4.0/). Funded by SCOAP[3].

problem [7,59]. For any $q$-form field in the bulk without topological obstructions, there is the well-known Hodge duality, namely, a bulk $q$-form field $X_{[q]}$ is dual to a bulk $(p-q)$-form field $X_{[p-q]}$ [37]. So there is a question how to make sure the $q$-form and its dual fields are simultaneously localized on the brane. That is why we proposed a new localization mechanism for the $q$-form field in Ref. [60] different from those in Refs. [37,61].

The line element is assumed as

$$ds^2 = e^{2A(z)}\left(\hat{g}_{\mu\nu}(x^\lambda)dx^\mu dx^\nu + dz^2\right), \quad (2)$$

where $A(z)$ is the warp factor depending only on the extra dimension coordinate $z$, and $\hat{g}_{\mu\nu}(x^\lambda)$ is the induced metric on the brane. In Ref. [61], based on the following ansatz for the zero mode:

$$X_{\mu_1\mu_2\cdots\mu_q}(x_\mu, z) = \hat{X}^{(n)}_{\mu_1\mu_2\cdots\mu_q}(x^\mu), \quad (3a)$$

$$X_{\mu_1\mu_2\cdots\mu_{q-1}z}(x_\mu, z) = 0, \quad (3b)$$

where the $\hat{\ }$ indicates the effective quantities on the brane, it was concluded that the zero (massless) mode of a massless $q$-form field with $q \geq (p-1)/2$ in the bulk cannot be localized on the RS $p$-brane, which indicates that only the zero mode of the scalar (the 0-form field) is trapped to the brane in the case of $p=3$ (while the zero mode of the 3-from field cannot be localized). This result conflicts with the well-known Hodge duality in the bulk without topological obstructions [62].

In order to solve the above contradiction, Duff and Liu suggested to keep the ansatz (3) for $q < (p-1)/2$ but modify it for $q \geq (p-1)/2$ as [37]

$$X_{\mu_1\mu_2\cdots\mu_q}(x_\mu, z) = 0, \quad (4a)$$

$$X_{\mu_1\mu_2\cdots\mu_{q-1}z}(x_\mu, z) = \hat{X}^{(n)}_{\mu_1\mu_2\cdots\mu_{q-1}}(x^\mu). \quad (4b)$$

With this new KK decomposition, Hodge duality in the bulk implies Hodge duality on the brane for the zero modes. However, Hodge duality on the brane for other KK modes cannot be obtained.

In order to further complete the investigation of localization and Hodge duality of $q$-form fields on the $p$-brane, we recently proposed a new general KK decomposition for a bulk $q$-form field without choosing a gauge in advance: [60]

$$X_{\mu_1\mu_2\cdots\mu_q}(x_\mu, z) = \sum_n \hat{X}^{(n)}_{\mu_1\mu_2\cdots\mu_q}(x^\mu)\, U_1^{(n)}(z)\, e^{(2q-p)A/2}, \quad (5a)$$

$$X_{\mu_1\mu_2\cdots\mu_{q-1}z}(x_\mu, z) = \sum_n \hat{X}^{(n)}_{\mu_1\mu_2\cdots\mu_{q-1}}(x^\mu)\, U_2^{(n)}(z)\, e^{(2q-p)A/2}, \quad (5b)$$

where the index $n$ marks different KK modes. There are two types of KK modes, the $q$-form modes $\hat{X}^{(n)}_{[q]}$ and the $(q-1)$-form ones $\hat{X}^{(n)}_{[q-1]}$. The functions $U_1^{(n)}(z)$ and $U_2^{(n)}(z)$ give respectively the distributions of the two types of KK modes along the extra dimension. They satisfy two non-independent Schrödinger-like equations [60], which will lead to some interesting results about the localization of these KK modes if we consider explicit brane models. With this KK decomposition new results for Hodge dualities were discovered. It was found that the Hodge duality in the bulk naturally results in two dualities on the brane. One is the duality between the zero modes $\hat{X}^{(0)}_{[q]}$ and $\hat{X}^{(0)}_{[p-q-1]}$, or between $\hat{X}^{(0)}_{[q-1]}$ and $\hat{X}^{(0)}_{[p-q]}$. The other is between two group KK modes, $(\hat{X}^{(n)}_{[q]}, \hat{X}^{(n)}_{[q-1]})$ and $(\hat{X}^{(n)}_{[p-q]}, \hat{X}^{(n)}_{[p-q-1]})$, where $\hat{X}^{(n)}_{[q]}$ and $\hat{X}^{(n)}_{[p-q]}$ are massive while $\hat{X}^{(n)}_{[q-1]}$ and $\hat{X}^{(n)}_{[p-q-1]}$ are massless. The effective field theories on the brane for these dual KK modes are physically equivalent [60].

In this paper, we would like to investigate localization and mass spectra for bulk $q$-form fields in two typical brane models by following the general KK decomposition (5). We first consider a $p$-brane model with a finite extra dimension [45]. Using numerical method we will find that for a bulk 3-form field, there are some massive bound 3-form KK modes $\hat{X}^{(n)}_{[3]}$ and a series of massless bound 2-form ones $\hat{X}^{(n)}_{[2]}$ with different extra dimensional configurations $U_2^{(n)}$ on the 3-brane. These 2-form KK modes are different from those got from a bulk 2-form field, because they come from the 2-form parts of a bulk 3-form field, as we can see from (5b). Besides, there are continuous unbound KK modes. Then we move to a degenerated 3-Bloch brane model [48]. For the bulk 3-form field, some resonant massive 3-form modes and the corresponding massless resonant 2-form ones appear. The $n$-level ($n > 0$) bound/resonant massive 3-form KK mode always has opposite parity with the $n$-level ($n > 0$) massless 2-form one.

This paper is organized as follows. We first review the new localization mechanism for a bulk $q$-form field on codimension-one $p$-branes in Sec. 2. Then we investigate the bound KK modes and the resonant ones for a bulk 3-form field in different $p$-brane models in Sec. 3. Finally, a brief conclusion is given in Sec. 4.

## 2. Review of the new localization mechanism

In this section, we briefly review the new localization mechanism for a bulk $q$-form field on codimension-one $p$-branes proposed in Ref. [60]. The action and line element are given in Eqs. (1) and (2), respectively. Substituting the general KK decomposition (5) into the action (1) for the $q$-form field, we get [60]

$$S_{\text{eff}} = -\frac{1}{2(q+1)!}\sum_{n,n'}\int d^{p+1}x\sqrt{-\hat{g}}\left[I^{(1)}_{nn'}\hat{Y}^{\mu_1\cdots\mu_{q+1}}_{(n)}\hat{Y}^{(n')}_{\mu_1\cdots\mu_{q+1}}\right.$$
$$\left.+ I^{(2)}_{nn'}\hat{Y}^{\mu_1\cdots\mu_q}_{(n)}\hat{Y}^{(n')}_{\mu_1\cdots\mu_q} + I^{(3)}_{nn'}\hat{X}^{\mu_1\cdots\mu_q}_{(n)}\hat{X}^{(n')}_{\mu_1\cdots\mu_q}\right.$$
$$\left.+ 2I^{(4)}_{nn'}\hat{Y}^{\mu_1\cdots\mu_q}_{(n)}\hat{X}^{(n')}_{\mu_1\cdots\mu_q}\right], \quad (6)$$

where we have assumed that $U_{1,2}^{(n)}(z)$ satisfy the following orthonormality conditions:

$$I^{(1)}_{nn'} \equiv \int dz\, U_1^{(n)} U_1^{(n')} = \delta_{nn'}, \quad (7a)$$

$$I^{(2)}_{nn'} \equiv \frac{q^2}{q+1}\int dz\, U_2^{(n)} U_2^{(n')} = (q+1)\,\delta_{nn'}, \quad (7b)$$

and $I^{(3,4)}_{nn'}$ are defined as

$$I^{(3)}_{nn'} \equiv \frac{1}{q+1}\int dz\, e^{(p-2q)A}\partial_z(U_1^{(n)}e^{(2q-p)A/2})\partial_z(U_1^{(n')}e^{(2q-p)A/2}), \quad (8)$$

$$I^{(4)}_{nn'} \equiv \frac{q}{q+1}\int dz\, e^{(p-2q)A/2} U_2^{(n)}\,\partial_z(U_1^{(n')}e^{(2q-p)A/2}). \quad (9)$$

From the KK decomposition, we see that there will be two types of KK modes for the bulk $q$-form field, i.e., the $q$-form KK modes $\hat{X}^{(n)}_{[q]}$ and the $(q-1)$-form ones $\hat{X}^{(n)}_{[q-1]}$. Further with the effective action (6), it is known that if the orthonormality conditions (7a) and (7b) can be satisfied, these two types of KK modes will be localized on the brane, and they couple with each other (in fact they can be decoupled if we choose a suitable gauge [60]). We also note that the $(q-1)$-form KK modes are always massless, while the $q$-form ones may be massless or massive.

We can introduce a mass parameter $m_n$ for the massive $q$-form KK modes for simplicity, i.e., $I^{(3)}_{nn'} = m_n\,\delta_{nn'}$. For different massive $q$-form KK modes, they surely have different masses. And in this sense, all the $(q-1)$-form KK modes seem to be the same on the brane, as they are all massless. However, in a higher-dimensional perspective, the KK modes have different distributions along the extra dimension. If we consider a thick brane, whose energy density also distributes along the extra dimension, there will be different probabilities for detecting the KK modes on different positions of the thick brane. Therefore, the massless $(q-1)$-form KK modes are in fact different along the extra dimension, which are described by the functions $U_2^{(n)}(z)$.

In order to deeply analyze the properties of these KK modes for the bulk $q$-form field, we need the equations they satisfy. To this end, we can derive the equations of motion for the KK modes from the effective action (6). On the other hand, we can substitute the KK decomposition (5) into the equations of motion for the bulk $q$-form field, and get another group of equations of motion. Comparing the two groups of equations, we finally find that the KK modes satisfy the following coupled equations [60]:

$$\partial_z U_2^{(n)}(z) + \frac{p-2q}{2} A'(z) U_2^{(n)}(z) = -\frac{q+1}{q} m_n U_1^{(n)}(z), \quad (10a)$$

$$\partial_z U_1^{(n)}(z) - \frac{p-2q}{2} A'(z) U_1^{(n)}(z) = +\frac{q}{q+1} m_n U_2^{(n)}(z), \quad (10b)$$

which can also be written as two non-independent Schrödinger-like equations:

$$\left[-\partial_z^2 + V_{q,1}(z)\right] U_1^{(n)}(z) = m_n^2 U_1^{(n)}(z), \quad (11a)$$

$$\left[-\partial_z^2 + V_{q,2}(z)\right] U_2^{(n)}(z) = m_n^2 U_2^{(n)}(z), \quad (11b)$$

where the effective potentials are given by

$$V_{q,1}(z) = \frac{(p-2q)^2}{4} A'^2(z) + \frac{p-2q}{2} A''(z), \quad (12a)$$

$$V_{q,2}(z) = \frac{(p-2q)^2}{4} A'^2(z) - \frac{p-2q}{2} A''(z). \quad (12b)$$

Then with a given brane solution, we can use the equations (11a) and (11b) to calculate the mass spectrum $m_n$ and the wave functions $U_1^{(n)}(z)$ and $U_2^{(n)}(z)$ for the $q$- and $(q-1)$-form KK modes, respectively. And the results are different from the former works [45,63,64], where only the Schrödinger-like equation (11a) was found.

Not all the KK modes solved from (11a) or (11b) can be localized on the brane, since the orthonormality conditions (7a) or (7b) cannot be satisfied automatically. In this paper, we are interested in the bound and resonant KK modes, which are the modes that can be localized and quasi-localized on the brane, respectively. The bound KK modes can be localized on the brane forever, while the resonances have finite lifetimes on the brane. Which types of KK modes will appear is determined by the brane solution. In the following discussion, we will consider two different $p$-brane models, i.e., a $p$-brane with a finite extra dimension and a degenerated 3-Bloch brane with an infinite extra dimension, and investigate these KK modes for the bulk $q$-form field under the above new localization mechanism.

## 3. New results under the localization mechanism

We have known that under the general KK decomposition (5) there will be some $q$-form KK modes and $(q-1)$-form ones for a bulk $q$-form field. It is also interesting to note that under the localization mechanism reviewed in previous section, if the orthonormality condition (7a) for a $n$-level massive $q$-form KK mode is satisfied, which will lead to the localization of this mode on the brane, the orthonormality condition (7b) will also be satisfied. This can be derived from Eqs. (10b) and (11a):

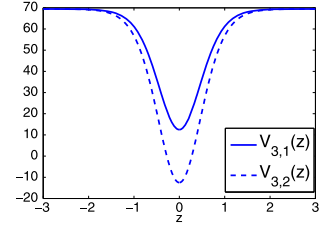

**Fig. 1.** The shapes of the effective potentials $V_{3,1}(z)$ and $V_{3,2}(z)$ for the warp factor (16) with $p=q=3$, $v=5$, $a=1$.

$$I_{nn'}^{(2)} = \frac{q+1}{m_n^2} \int dz \left(\partial_z U_1^{(n)}(z) - \frac{p-2q}{2} A'(z) U_1^{(n)}(z)\right)$$
$$\quad \left(\partial_z U_1^{(n')}(z) - \frac{p-2q}{2} A'(z) U_1^{(n')}(z)\right)$$
$$= (q+1) \int dz\, U_1^{(n)}(z) U_1^{(n')}(z). \quad (13)$$

This means that there will be a localized $n$-level massless $(q-1)$-form KK mode accompanying with a localized $n$-level massive $q$-form one. However, for the 0-level KK modes, the $q$-form and the $(q-1)$-form decouple from each other, and at most only one of them can be localized. These new results can also be understood under supersymmetric mechanism [60]. So in the following we will expect to find some massless bound/resonant $(q-1)$-form KK modes, not only the $q$-form ones as that appeared in the former works.

### 3.1. Bound KK modes

In Ref. [45], the authors gave a $p$-brane solution with a finite extra dimension, which is constructed by two scalar fields $\phi$ and $\pi$. The action of this system is

$$S = \int d^D x \sqrt{-g} \left[\frac{1}{2\kappa_D^2} R - \frac{1}{2}(\partial\phi)^2 - \frac{1}{2}(\partial\pi)^2 - V(\phi,\pi)\right]. \quad (14)$$

Then, with the conformally flat metric (2) and using the superpotential method, the solution for the brane system was found as [45]

$$\phi = v \tanh(az), \qquad \pi = \sqrt{p}\, A, \quad (15)$$

$$A = -\frac{v^2}{3p}\left[\ln\cosh^2(az) + \frac{1}{2}\tanh^2(az)\right], \quad (16)$$

where $a$ and $v$ are parameters.

With the warp factor (16), the effective potentials $V_{q,1}(z)$ and $V_{q,2}(z)$ in (12) for the $q$- and $(q-1)$-form KK modes can be got, for which the asymptotic behaviors are

$$V_{q,1}(0) = -V_{q,2}(0) = -\frac{a^2 v^2 (p-2q)}{2p}, \quad (17)$$

$$V_{q,1}(z\to\infty) = V_{q,2}(z\to\infty) \to \frac{a^2 v^4 (p-2q)^2}{9p^2}. \quad (18)$$

It is clear that $V_{q,1}(z)$ and $V_{q,2}(z)$ have opposite values at $z=0$, but the same value at infinity, and they are Pöschl–Teller-like (PT-like) potentials with $v^2 > \frac{9p}{2|p-2q|}$. Thus there always exist some bound KK modes for any $q$-form field.

Here we take a bulk 3-form field as an example to numerically solve Eqs. (11a) and (11b) and get the bound KK modes. The shapes of the effective potentials for the bulk 3-form field are plotted in Fig. 1 for $p=q=3$, $v=5$, $a=1$.

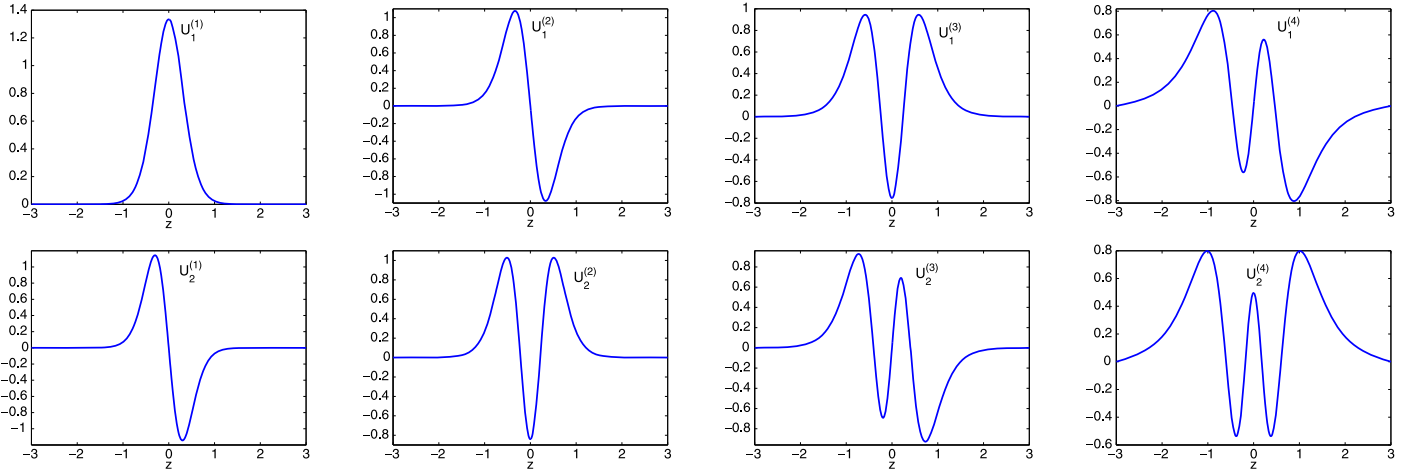

**Fig. 2.** The shapes of $U_1^{(n)}(z)$ and $U_2^{(n)}(z)$ for the bound KK modes of a 3-form field on a 3-brane with a finite extra dimension with $p = q = 3$, $v = 5$, $a = 1$.

The mass spectrum for the bound 3-form KK modes $U_1^{(n)}$ is solved from the Schrödinger-like equation (11a) as

$$m_n^2 = (22.958, 41.725, 56.041, 65.513), \quad (19)$$

which shows that there exist four massive bound 3-form KK modes. Note that there is no massless mode, because the potential is positive.

For the massless 2-form KK modes, there exists a bound 0-level mode (the ground state $U_1^{(0)}$) besides four excited bound modes corresponding to the same mass spectrum as in (19). All the five KK modes are normalizable (i.e., $\int dz \, (U_2^{(n)})^2 < \infty$) since they exponentially decay to zero at the boundary of extra dimension. Note that although we get a mass spectrum $m_n^2 = (0, 22.958, 41.725, 56.041, 65.513)$, all the 2-form KK modes are massless and the mass parameters come from the couplings with the 3-form KK modes.

We plot the shapes of $U_1^{(n)}(z)$ and $U_2^{(n)}(z)$ for the massive bound 3-form KK modes and the corresponding massless 2-form ones in Fig. 2. It is clear that the parities of the $U_1^{(n)}(z)$ and $U_2^{(n)}(z)$ ($n \geq 0$) are opposite. The wave function $U_1^{(1)}(z)$ for the 1-level massive bound 3-form KK mode is even, and the corresponding $U_2^{(1)}(z)$ for the 1-level massless 2-form mode is odd. This is because there is a 0-level localized massless 2-form mode solved from (11b).

According to above analyse, in this background with $v^2 > \frac{9p}{2|p-2q|}$, if we consider a bulk 1-form (vector) field, there must exist some bound 1-form KK modes accompanying with some bound 0-form (scalar) ones. It seems similar to the Stueckelberg action for a massive vector field. The difference is that for the Stueckelberg mechanism the scalar field is responsible for the mass of the vector field, but here it suggests that the masses of vector fields are because of the geometry of the brane background, and the corresponding scalar fields are just one part of the higher dimensional massless vector field.

### 3.2. Resonant KK modes

For the massive $q$-form resonances, they in fact share some similarities with the bounded ones. We know that, in some braneworld models, there are no bound KK modes that can be localized on the brane; but some discrete modes can be quasi-localized on the brane (these modes are called KK resonances). This means that within a finite time the four-dimensional KK particles on the brane can be obtained, but they will propagate into the bulk at last. For these KK resonances the relationship (13) is also satisfied, which indicates that if there is a massive $q$-form resonance, there will be an accompanying massless $(q-1)$-form one. This conclusion is similar to the case of the bound KK modes, which will be discussed in this section.

To investigate the resonant KK modes, we consider a degenerated 3-Bloch brane, which is also constructed by two scalars. For this brane system, the line element is assumed as

$$ds^2 = e^{2A(y)}\eta_{\mu\nu}dx^\mu dx^\nu + dy^2, \quad (20)$$

where $y$ stands for the extra dimension. Here, we only list the solution of the warp factor [17,48,65]:

$$e^{2A(y)} = \left(\frac{c_0 - u}{c_0 - u\cosh(2bdy)}\right)^{4v^2/9}$$
$$\times \exp\left[\frac{4uv^2}{9}\left(\frac{u - c_0\cosh(2bdy)}{(c_0 - u\cosh(2bdy))^2} - \frac{1}{u - c_0}\right)\right], \quad (21)$$

where $b, d, u, c_0$ are constants, and $u = \sqrt{c_0^2 - 4}$. For $u \to 0$, there will appear a double brane, which is called a degenerated Bloch brane. For convenience, we redefine $u = 4e^{-\delta_0}$ with $\delta_0 \geq 1$ [48].

After a coordinate transformation $dz = e^{-A}dy$, the effective potentials (12a) and (12b) for the KK modes of the 3-form field can be rewritten in $y$-coordinate as

$$V(z(y))_{q,1} = e^{2A}\left[\frac{(3-2q)^2 + 2(3-2q)}{4}(\partial_y A)^2 + \frac{3-2q}{2}\partial_y^2 A\right], \quad (22)$$

$$V(z(y))_{q,2} = e^{2A}\left[\frac{(3-2q)^2 + 2(2q-3)}{4}(\partial_y A)^2 + \frac{2q-3}{2}\partial_y^2 A\right]. \quad (23)$$

We also consider a bulk 3-form field. Then with the solution of the warp factor (21), the shapes of the effective potentials (22) and (23) in this brane model are shown in Fig. 3(a) with $d = 1, b = 6, \delta_0 = 25$. We see that for both potentials there are potential barriers, which will lead to some resonances.

We can use the method in Ref. [44] to find resonances. We define the relative probability for the KK mode within a narrow range $-z_b \leq z \leq z_b$ around the thick brane location [44]:

$$P_i = \frac{\int_{-z_b}^{z_b} |U_i^{(n)}|^2 dz}{\int_{-z_{max}}^{z_{max}} |U_i^{(n)}|^2 dz}, \quad (i = 1, 2) \quad (24)$$

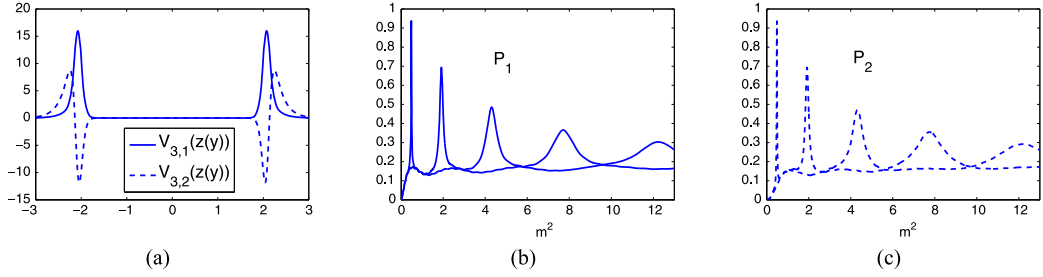

**Fig. 3.** The shapes of the effective potentials $V_{3,1}(z(y))$ and $V_{3,2}(z(y))$, and the relative probabilities $P_1$ and $P_2$ for a 3-form field in a degenerated 3-Bloch brane with $d = 1, b = 6, \delta_0 = 25$.

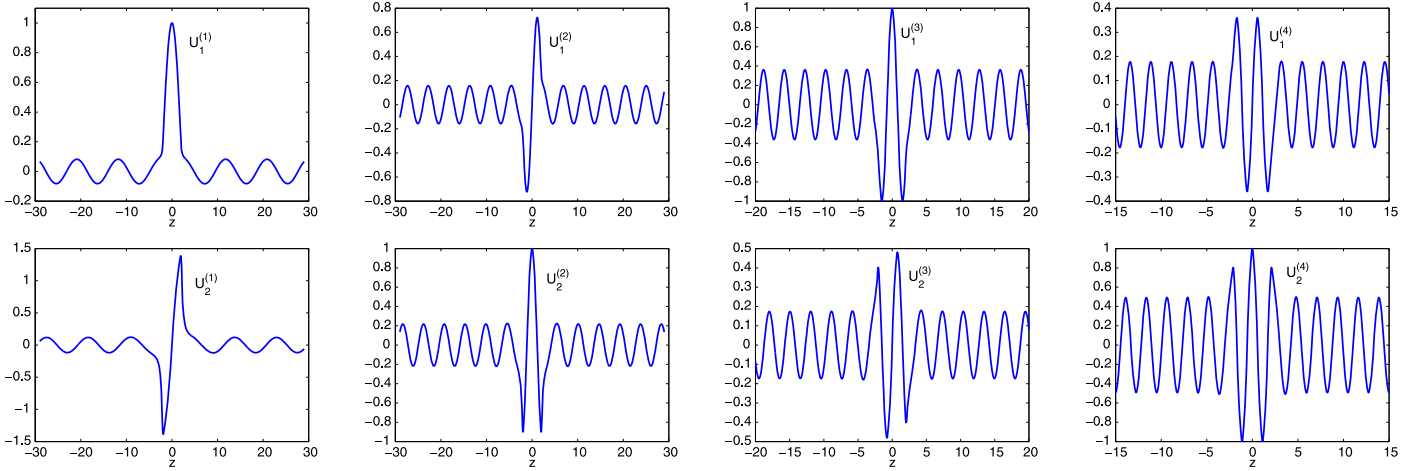

**Fig. 4.** The shapes of $U_1^{(n)}(z)$ and $U_2^{(n)}(z)$ for the resonant KK modes of a bulk 3-form field in a degenerated 3-Bloch brane with $d = 1, b = 6, \delta_0 = 25$.

**Table 1**
The mass $m^2$, width $\delta m$, and lifetime $\tau$ for resonances of a 3-form field in a degenerated 3-Bloch brane with $d = 1, b = 6, \delta_0 = 25$.

| n | $m^2$ | m | $\delta m$ | $\tau$ |
|---|-------|-------|-------|--------|
| 1 | 0.483 | 0.695 | 0.027 | 37.551 |
| 2 | 1.914 | 1.384 | 0.091 | 10.948 |
| 3 | 4.306 | 2.075 | 0.227 | 4.404  |
| 4 | 7.711 | 2.777 | 0.603 | 1.658  |

where we choose $z_{max} = 10 z_b$. For different KK modes the $U_1^{(n)}$ and $U_2^{(n)}$ can be solved respectively from Eqs. (11a) and (11b) with two kinds of initial conditions: $U_i^{(n)}(z) = 0, U_i'^{(n)}(z) = h_i$ for odd KK modes, and $U_i(z) = k_i, U_i'(z) = 0$ for even ones, where $h_i$ and $k_i$ are arbitrary nonvanishing constants but they can be determined by some normalization conditions. Then a large relative probability $P_i$ will indicate the existence of a resonance.

There are also expected to be some massive 3-form resonant KK modes, and some accompanying massless 2-form ones. For the former, they can be numerically calculated from the Schrödinger-like equation (11a) or (11b). The mass spectrum and their lifetime $\tau$ [1] of the 3-form resonances are listed in Table 1. The pictures for $P_1$ and $P_2$ are shown in Figs. 3(b) and 3(c). From Table 1 we see that the heavier resonances will have shorter lifetime on the brane.

Then the massless 2-form resonances can be found from the Schrödinger-like equation (11b). We plot the wave functions $U_1^{(n)}(z), U_2^{(n)}(z)$ in Fig. 4, from which one can see that the wave function for 1-level massive 3-form resonance $U_1^{(1)}(z)$ is even, but the corresponding 1-level massless 2-form one $U_2^{(1)}(z)$ is odd.

---
[1] The $\tau$ is defined as $\tau \sim (\delta m)^{-1}$ with $\delta$ the width in mass at half maximum of each peak in Fig. 3(b) or 3(c) [44].

This is because there is a 0-level massless bound 2-form mode $U_1^{(0)}(z) \propto e^{-3A/2}$, which is an even function. So that other $n$-level 3-form modes also have opposite parities with the $n$-level 2-form ones.

Through our above discussions about the localization of the q-form field, it is interesting to note that there are some similarities to the fermions localization [44,66]. Firstly, the two Schrodinger-like equations for the $q$-form and $(q - 1)$-form KK modes can be written by analoguos supersymmetric operators just like those for left-handed and right-handed fermions in braneworld models detailed in Ref. [66]:

$$\mathcal{Q}\mathcal{Q}^\dagger U_1^{(n)}(z) = m_n^2 U_1^{(n)}(z), \qquad (25)$$

$$\mathcal{Q}^\dagger \mathcal{Q} U_2^{(n)}(z) = m_n^2 U_2^{(n)}(z), \qquad (26)$$

with the operator $\mathcal{Q}$ given by $\mathcal{Q} = \partial_z + \frac{p-2q}{2} A'(z)$. This ensures the absence of negative eigenvalues. Secondly, the eigenfunctions for the $q$-form and $(q - 1)$-form KK modes have opposite parities as shown in Figs. 2 and 4, which was also verified for the left-handed and right-handed fermions in Ref. [44]. The difference for the $q$-form field and fermion localization is that for the fermion the left- and right-handed fermions share the same mass spectra except for the zero modes; but for the $q$-form field, the $q$-form KK modes are massive while the $(q-1)$-form KK ones are all massless.

## 4. Conclusion

In this work, using a localization mechanism for a bulk $q$-form field without any prior gauge choice, we investigated the KK modes of the field on $p$-branes with codimension one. It was found that there may exist some $q$-form and $(q - 1)$-form KK modes for the bulk $q$-form field. These two types of KK modes satisfy two non-independent Schrödinger-like equations, which indicates that at most one of the 0-level $q$-form and $(q-1)$-form modes can be localized on the brane, but a localized $n$-level ($n \neq 0$) massive $q$-form mode will appear accompanying with a localized $n$-level massless $(q-1)$-form mode. These results are new and different from the former works. Note that, with this localization mechanism, the Hodge dualities for both zero modes and nonzero modes on the brane can be naturally realized.

To see these new results clearly, we investigated the KK modes of a bulk $q$-form field in different $p$-brane models. On a 3-brane with a finite extra dimension, through some numerical calculations for a bulk 3-form field, we found some massive 3-form bound KK modes and the corresponding massless 2-form bound ones. While on a degenerated 3-Bloch brane with an infinite extra dimension, some massive 3-form resonances and massless 2-form ones are obtained. In both models, the wave functions $U_1^{(n)}(z)$ for the 3-form KK modes and $U_2^{(n)}(z)$ for the 2-form KK modes have opposite parities, which is a result of the coupled equations (10).

## Acknowledgements

We would like to thank Prof. Bob Holdom for his useful discussion for this work. This work was supported by the National Natural Science Foundation of China (Grant Nos. 11375075, 11405121, and 11522541). C.-E. Fu was also supported by the scholarship granted by the Chinese Scholarship Council (CSC). Q.-Y. Xie was supported by the Fundamental Research Funds for the Central Universities (Grant No. lzujbky-2015-107).